# Coherent terahertz needle radiation emission mechanism from nanostructured ferromagnetic metal films excited by femtosecond laser pulses

**Anatoly Gladun, Vladimir Leiman, and Aleksey Arsenin**[*]

*Department of General Physics, Moscow Institute of Physics and Technology (State University), 9 Institutskii Lane, Dolgoprudny, 141700, Russia*
[*]*Corresponding author: a.arsenin@mail.ru*

**Abstract:** A coherent terahertz needle radiation emission mechanism is proposed. This mechanism can be realized in the experiments on femtosecond laser pulses interacting with surface of nanostructured ferromagnetic metal film. The proposed mechanism is based on exciting of coupled coherent phonon-magnon modes on a nanostructured metal surface by laser femtosecond pulse. The value of the Poynting vector for the terahertz radiation is estimated.

**References and links**

1. M. Tonouchi, "Cutting-edge terahertz technology," Nature Photonics **1**, 97-105 (2007).
2. P.H. Siegel, "Terahertz technology," IEEE Trans. Microwave Theory Techn. **MTT-50**, 910–928 (2002).
3. U. Bovensiepen, "Coherent and incoherent excitations of the Gd(0001) surface on ultrafast timescales," J. Phys.: Condens. Matter **19**, 083201 (2007).
4. M. Vomir, L.H.F. Andrade, E. Beaurepaire, M. Albrecht, and J.-Y. Bigot, "Ultrafast magnetization dynamics investigated in real space (invited)," J. Appl. Phys. **99**, 08A501 (2006).
5. A. Laraoui, J. Vénuat, V. Halté, M. Albrecht, E. Beaurepaire, and J.-Y. Bigot, "Study of individual ferromagnetic disks with femtosecond optical pulses," J. Appl. Phys. **101**, 09C105 (2007).
6. A. Melnikov, I. Radu, U. Bovensiepen, O. Krupin, K. Starke, E. Matthias, and M. Wolf, "Coherent optical phonons and parametrically coupled magnons induced by femtosecond laser excitation of the Gd(0001) surface," Phys. Rev. Lett. **91**, 227403 (2003).
7. E. Beaurepaire, G.M. Turner, S.M. Harrel, M.C. Beard, J.-Y. Bigot, and C.A. Schmuttenmaer, "Coherent terahertz emission from ferromagnetic films excited by femtosecond laser pulses," Appl. Phys. Lett. **84**, 3465–3467 (2004).

Terahertz (THz) radiation lies in the frequency gap between infrared and microwaves, and its typical frequencies range from 100 GHz to 30 THz. Today research into terahertz technology is attracts attention throughout the world. Recent technological innovation in photonic and nanotechnology is stimulating THz research to be applied in very wide range of applications [1, 2]: information and communication technologies; biology and medical sciences; non-destructive evaluation; homeland security; quality control of food and agricultural products; environmental monitoring and ultrafast computing; etc.

In order to make the mentioned applications possible one needs higher-power THz sources, more sensitive THz sensors, and more functional devices and materials. Thus, finding effective THz sources is an actual scientific problem.

It is well known that the conventional demagnetization processes (spin precession, magnetic domain motion and rotation) are governed mainly by spin-lattice, magnetic dipole-dipole, Zeeman, spin-spin interactions. It occurs on a timescale of nanoseconds. Fortunately and unprecedentedly, recent experimental investigations have evidenced much faster magnetization dynamics with character time of a few femtoseconds: femtomagnetism "terra incognita". The question is the excitation of a ferromagnetic solid or film by a femtosecond laser pulses [3–5].

In this temporal region the classical Landau-Lifshitz-Gilbert equation fails and theoretical description must be based on quantum mechanics. This novel spin dynamics has not been well-understood yet.

Except the induced electron and spin dynamics the laser pulses can also trigger coherent lattice dynamics, in particular generate the optical phonon mode, if the ferromagnet has two atoms in a primitive unit cell. However such coherent optical phonons were not expected to occur in metals due to the effective screening of the spatial electron redistribution that takes place on a timescale equal with the inverse of the plasma frequency (attosecond timescale).

Nevertheless the work [6] e.g. presents the novel phenomenon of coupled coherent phonon-magnon mode measured on the ferromagnetic Gd (0001) surface by means of time-resolved magnetization-induced second-harmonic generation. The coupled quasiparticle constitutes itself in new type of phonon-magnon interaction whose coupling is mediated by the exchanged interaction. The resulting magnetization dynamics is considered as optical magnon wave packets coupled to the phonon. It is clear that a standing wave of such coupled mode on the sample of a finite size is possible.

The energy flux in a standing wave equals zero. Every section of a sample of the length $\lambda/4$ (where $\lambda$ is the wavelength) doesn't exchange of energy with the neighboring sections. Its energy is constant. In every such section the potential energy is transforming to the kinetic energy twice during the period. This means that the analogy between a standing wave and self-oscillations of an oscillator does take place. The corresponding frequencies are the natural or resonant frequencies of the sample. Thereby the magnetic moment of the sample has sinusoidal time dependence (if one neglects the damping).

In the letter [7] it is shown that the laser induced ultrafast demagnetization of ferromagnetic films results in the emission of a terahertz electromagnetic pulse. The radiated electric field $E(t)$ is explained by Maxwell equations (radiation from a time dependent magnetic dipole), and is expected to be proportional to the second time derivative of the magnetization $d^2M/dt^2$, as was measured in the far field. We suppose that the emission of a terahertz electromagnetic pulse here is conditioned by the phonon-magnon modes on a metal surface.

It's reasonable to suggest the mechanism which can essentially intensify this effect.

In this paper we exploit the idea that the femtosecond laser pulse induces the magnetic dipole wave on a nanostructured ferromagnetic metal film due to an excitation of the coupled magnon-phonon standing mode in every elementary cell of the structured surface (fig. 1). The standing magnetic dipole wave is terahertz needle-like radiation (similar to the Einstein's needle-radiation) emission source.

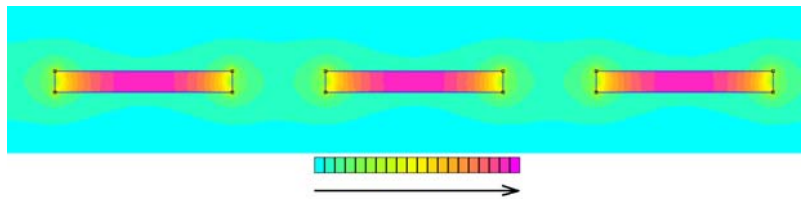

Fig. 1. A physical model of the relevant nanostructured ferromagnetic film surface. Arrow notes increasing flux density.

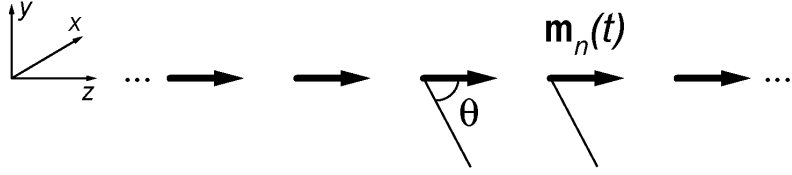

Fig. 2. Periodic structure, consisting of point magnetic moments (one-dimensional representation of a grating with the period $\Lambda$).

We clarify the radiation mechanism essence via a simple model. Consider an infinite periodic structure (grating) with the period $\Lambda$, consisting of identical point magnetic moments $\mathbf{m}_n(t)$, arranged along the $z$ axis (fig. 2). The expression for the magnetization vector $\mathbf{M}$ of the structure can be written in the form

$$\mathbf{M} = \sum_{n=-\infty}^{\infty} \mathbf{m}_n(t)\delta(z-n\Lambda)\delta(y),$$

$$\frac{\partial}{\partial x} = 0.$$

Here symbol $\delta$ designates Dirac delta-function. Consider the emission of radiation from this system at large distances (in wave zone). The angle $\theta$ in fig. 2 determines the direction to the observation point. Let us introduce the retarded Hertz vector

$$\mathbf{\Pi} = \frac{\mu_0}{4\pi} \int_V \frac{\mathbf{M}(\mathbf{r}', t-\frac{r}{c})}{r} d\mathbf{r}',$$

where $\mathbf{r}'$ is the radius vector characterising the dipole position; $r$ is the modulus of the radius vector drawn from the dipole to the observation point; $c$ is the speed of light in vacuum and $\mu_0$ is the permeability of free space. The potentials $\varphi$ and $\mathbf{A}$ of the electromagnetic field are determined by the expressions

$$\varphi = 0, \quad \mathbf{A} = \mathrm{rot}\,\mathbf{\Pi}.$$

The vector $\mathbf{\Pi}$ satisfies the wave equation

$$\left(\nabla^2 - \frac{1}{c^2}\frac{\partial^2}{\partial t^2}\right)\mathbf{\Pi} = -\mu_0 \mathbf{M}.$$

For the radiating system under study, we have

$$\mathbf{\Pi} = \sum_{n=-\infty}^{\infty} \frac{1}{r_n}\mathbf{m}_n\left(t-\frac{r_n}{c}\right) = \frac{1}{r}\sum_{n=-\infty}^{\infty} \mathbf{m}_n\left(t-\frac{r_n}{c}\right), \qquad (1)$$

where $r_n$ is the modulus of the radius vector drawn from the $n$-th dipole to the observation point. The latter equality in (1) is justified because $1/r_n$ slowly varies at large distances from the periodic structure. The moment $m_n$ has the components {0, 0, m} in the Cartesian coordinates. We assume that

$$\boldsymbol{m}_n = \boldsymbol{a}\exp[-i(\omega t + k_g \xi)], \qquad (2)$$

where $\xi = n\Lambda$ $(n = 0, \pm 1, \pm 2,...)$, $\boldsymbol{a}$ - a complex vector ($\boldsymbol{a} = |\boldsymbol{a}|\exp(i\varphi)$), $k_g$ is the modulus of the wave vector of a wave in the grating. This means that the wave, which length $\lambda_g = 2\pi/k_g$, propagates in the periodic structure; $\omega = \omega(k_g)$ is the dispersion relation. By substituting (2) into (1), we find

$$\Pi = \frac{a\exp\left[-i\omega\left(t - \frac{r}{c}\right)\right]}{r} \sum_{n=-\infty}^{\infty} \exp\left[-in\Lambda\left(\frac{\omega}{c}\cos\theta + k_g\right)\right]. \qquad (3)$$

Where we took into account that $r_n = r - \xi\cos\theta$.

If the radiation wavelength $\lambda = 2\pi/\omega \gg \Lambda$ it is competent to substitute in (3) the summation by the integration since in this case it is possible to introduce a physically infinitely small interval $\Delta\xi$, that satisfies the condition

$$\lambda \gg \Delta\xi \gg \Lambda, \quad \Delta n = \frac{\Delta\xi}{\Lambda},$$

Here $\Delta n$ is the number of magnetic moments in this interval. By replacing summation in (3) by integration, we obtain:

$$\Pi = \frac{a\exp\left[-i\omega\left(t - \frac{r}{c}\right)\right]}{r} \cdot \frac{1}{\Lambda} \int_{-\infty}^{\infty} \exp\left[-i\left(\frac{\omega}{c}\cos\theta + k_g\right)\right] d\xi =$$

$$= \frac{a\exp\left[-i\omega\left(t - \frac{r}{c}\right)\right]}{r} \cdot \frac{2\pi}{\Lambda} \delta\left(\frac{\omega}{c}\cos\theta + k_g\right), \qquad (4)$$

which gives

$$\frac{\omega}{c}\cos\theta = -k_g. \qquad (5)$$

It follows from (5) that $\cos\theta < 0$ and $\frac{\omega}{k_g} = \frac{c}{\cos\theta}$, i.e. the phase velocity of the wave in the grating exceeds the speed of light in vacuum.

Consider now the standing waves radiation in the grating of length *l* in the wave zone (?). It is possible to consider the every element (the length $\Delta\xi \ll \lambda$) radiation summing fields from each separate magnetic moment by neglecting the radiation phase lag. In the wave zone the fields from a point magnetic moment that varies according to the harmonic law $\mathbf{m} = \mathbf{m}_0 \cos(\omega t - kr)$ are described by the following expressions (in the spherical coordinate system $r\varphi\theta$)

$$\mathbf{\Pi} = \frac{\mu_0}{4\pi} \frac{\mathbf{m}_0}{r} \exp\left[-i\omega\left(t - \frac{r}{c}\right)\right], \quad \omega = kc;$$

$$\mathbf{B} = \operatorname{Re}\{\mathbf{B}_0 \exp[-i(\omega t - kr)]\};$$

$$\mathbf{E} = \operatorname{Re}\{\mathbf{E}_0 \exp[-i(\omega t - kr)]\}; \quad (6)$$

$$\mathbf{B}_0 = \frac{\mu_0 \omega^2}{4\pi c^2 r}[\mathbf{e}_r, [\mathbf{e}_r, \mathbf{m}_0]];$$

$$\mathbf{E}_0 = c[\mathbf{B}_0, \mathbf{e}_r],$$

where $\mathbf{e}_r, \mathbf{e}_\varphi, \mathbf{e}_\theta$ are the basis vectors.

On the ground of (6) the total field of a set of $\Delta n$ magnetic moments at the length $\Delta \xi = \Lambda \cdot \Delta n$ is

$$dB_\theta = \frac{\mu_0 \omega^2 \sin\theta}{4\pi c^2 r} m_0(\xi) \cos(\omega t - kr) \frac{d\xi}{\Lambda},$$

where $r = r_0 - \xi\cos\theta$, $r_0$ is the distance measured from the grating centre to the observation point, $m_0(\xi) = m_0 \cos(k_g \xi - \varphi)$, $m_0 = 2|a|$, $B_\theta$ - is the scalar product $(\mathbf{B}\mathbf{e}_\theta)$.

By integrating over all the elements along the grating and taking into account the delay $\left(t \to t - \frac{r}{c}\right)$, we obtain

$$B_\theta = \frac{\mu_0 \omega^2}{4\pi c^2} \int_{-l/2}^{l/2} m_0(\xi) \frac{\sin\theta}{r} \cos(\omega t - kr_0 + k\xi\cos\theta) \frac{d\xi}{\Lambda}. \quad (7)$$

When calculation of (7) it is possible to factor out the expression $\sin\theta/r$ since the observation point is situated far from the grating

$$B_\theta = \frac{\mu_0 \omega^2 \sin\theta}{4\pi c^2 r} \int_{-l/2}^{l/2} m_0(\xi) \cos(\omega t - kr_0 + k\xi\cos\theta) \frac{d\xi}{\Lambda}. \quad (8)$$

It is possible to satisfy the boundary condition $m_0\left(\frac{l}{2}\right) = m_0\left(-\frac{l}{2}\right) = 0$ if we set $m_0(\xi) = m_0 \sin\frac{\pi s}{l}\xi$ at even $s$; $m_0(\xi) = m_0 \cos\frac{\pi q}{l}\xi$ at odd $q$. Here $s$ and $q$ are the whole numbers.

The calculation the integrals (8) in this case gives:

$$\int_{-l/2}^{l/2} \sin\frac{\pi s}{l}\xi \cos(k\xi\cos\theta + \omega t - kr_0)d\xi =$$

$$= \frac{\frac{\pi s}{l}}{\left(\frac{\pi s}{l}\right)^2 - (k\cos\theta)^2} \cdot 2\sin\left(\frac{kl}{2}\cos\theta\right)\cos(\omega t - kr_0);$$

$$\int_{-l/2}^{l/2} \cos\frac{\pi s}{l}\xi \cos(k\xi\cos\theta + \omega t - kr_0)d\xi =$$

$$= \frac{\frac{\pi s}{l}}{\left(\frac{\pi s}{l}\right)^2 - (k\cos\theta)^2} \cdot 2\cos\left(\frac{kl}{2}\cos\theta\right)\cos(\omega t - kr_0).$$

Finally, for even $s$ we obtain

$$B_\theta = \frac{\mu_0 \omega^2 \sin\theta}{2\pi c^2 r_0 \Lambda} \cdot \frac{\frac{\pi s}{l}}{\left(\frac{\pi s}{l}\right)^2 - (k\cos\theta)^2} \cdot \sin\left(\frac{kl}{2}\cos\theta\right)\cos(\omega t - kr_0);$$

$$E_\varphi = -cB_\theta.$$

Here $E_\varphi$ is equal to $(\mathbf{E}\mathbf{e}_\varphi)$. For odd $q$, a similar expression is obtained in which $\sin\left(\frac{kl}{2}\cos\theta\right)$ is replaced by $\cos\left(\frac{kl}{2}\cos\theta\right)$.

The value of the Pointing vector is appears to be

$$\frac{c}{\mu_0}B_\theta^2 = \frac{\mu_0 \omega^4 m_0^2 \sin^2\theta}{4\pi^4 c^3 r_0^2 s^2}\left(\frac{l}{\Lambda}\right)^2 \frac{\sin^2\left(\frac{kl}{2}\cos\theta\right)}{\left[1 - \left(\frac{k}{k_g}\cos\theta\right)^2\right]^2} \cos^2(\omega t - kr_0). \qquad (9)$$

Here $k = \frac{2\pi}{\lambda}$, $k_g = \frac{\pi s}{l}$, where the whole number $s$ is even. Thus we deal with the radiation known as needle radiation. It is obvious that $m_0 = \frac{\mu_s N}{l}\Lambda$, where $\mu_s$ is the saturation magnetization per atom, $N$ is the number of atoms in the ferromagnetic film. Therefore time averaged value of (9) equals to

$$\frac{c}{\mu_0}\overline{B_\theta^2} = \frac{2\mu_0 \nu^4 \mu_s^2 N^2}{c^3 r_0^2 s^2} \frac{\sin^2\left(\frac{kl}{2}\cos\theta\right)}{\left[1-\left(\frac{k}{k_g}\cos\theta\right)^2\right]^2}\sin^2\theta. \qquad (10)$$

where $\nu = \omega/2\pi$. The amount of energy emitted by the grating per unit time per solid angle element $d\Omega$ is $dI = \frac{c}{\mu_0}B_\theta^2 r_0^2 d\Omega$, where $d\Omega = 2\pi\sin\theta d\theta$.

Let estimate the numerical value of (10). From relation $\frac{kl}{2}\cos\theta = \pm\frac{\pi}{2}$ we find $\cos\theta = \pm\frac{\lambda}{2l}$. If for example $s = 2$, $\Lambda = 800$ nm, $\lambda = l = 0.3$ mm ($\nu = 1$ THz), we have $\theta = \pm 60°$, $k_g = \frac{2\pi}{l}$, $\frac{k}{k_g}\cos\theta = \frac{1}{2}$. The saturation magnetization $\mu_s$ for single-crystal Fe is equal to $3.7\mu_B$ atom$^{-1}$, where $\mu_B = \frac{e\hbar}{m} = 10^{-23}$ J/T is the Bohr magneton. Let the film dimensions be $3\cdot 10^{-4}$ m $\times$ $3\cdot 10^{-4}$ m $\times$ $5\cdot 10^{-8}$ m, the atoms concentration $n_a = 10^{29}$ m$^{-3}$, therefore $N = 4.5\cdot 10^{14}$ atoms; let also $r_0 = 1$ m. Then we have $\frac{c}{\mu_0}\overline{B_\theta^2} \approx 8.6$ W/m$^2$. For the distance $r_0 = 1$ cm: $\frac{c}{\mu_0}\overline{B_\theta^2} \approx 8.6\cdot 10^4$ W/m$^2$. This evaluation proves the availability of proposed terahertz radiation emission mechanism from nanostructured ferromagnetic films. Since this estimate is obviously to be too high, the suggested mechanism is suitable at least for researching of the femtomagnetism.

This work was supported by the Russian Foundation for Basic Research (Grant No. 08-02-00569) and grant for young scientists of the President of the Russian Federation No. MK-4082.2007.2.